\documentclass[10pt,nofootinbib,twocolumn,aps,prd]{revtex4-1}
\usepackage{amsmath,mathptmx,amssymb}
\usepackage{CJK}
\usepackage{graphicx,graphics,subfigure}
\usepackage{tabularx,ctable}
\usepackage[colorlinks=true,linkcolor=blue,citecolor=blue,urlcolor=blue]{hyperref}

\def\be{\begin{equation}}
\def\ee{\end{equation}}
\def\bea{\begin{eqnarray}}
\def\eea{\end{eqnarray}}
\def\ba{\begin{aligned}}
\def\ea{\end{aligned}}
\def\nn{\nonumber}
\def\p{\partial}

\def\cF{\mathcal{F}}
\def\cK{\mathcal{K}}

\begin{document}

\begin{CJK*}{GBK}{song}

\title{Topological classes of thermodynamics of the four-dimensional static accelerating black holes}

\author{Di Wu}
\email{wdcwnu@163.com}

\affiliation{School of Physics and Astronomy, China West Normal University,
Nanchong, Sichuan 637002, People's Republic of China}

\date{\today}

\begin{abstract}
In this paper, utilizing the generalized off shell Helmholtz free energy, we explore the topological
numbers of the four-dimensional static accelerating black hole and its AdS extension, as well as the
static charged accelerating black hole and its AdS extension. Our analysis reveals a profound and
significant impact of the acceleration parameter on the topological numbers associated with the static
black holes; and different values (nonzero) of the acceleration parameter do not
affect the topological numbers of the corresponding four-dimensional static accelerating black holes. In
addition, we demonstrate that the electric charge parameter has an important effect on the topological
number of the static neutral accelerating black holes, and the cosmological constant has a remarkable
influence on the topological number of the static accelerating black hole. Furthermore, it is interesting
to observe that the difference between the topological number of the asymptotically flat static accelerating
black hole and that of its corresponding asymptotically flat static nonaccelerating black hole is always unity,
and the difference between the topological number of the asymptotically AdS static accelerating black hole and
that of its corresponding asymptotically AdS static nonaccelerating black hole is always $-1$. This new
observation leads us to conjure that it might be valid also for other accelerating black holes. Of course,
this captivating conjecture requires empirical verification through comprehensive investigation into the
topological numbers of other accelerating black holes and their corresponding usual counterparts.
\end{abstract}

\maketitle
\end{CJK*}

\section{Introduction}
In the big family of four-dimensional black hole solutions in General Relativity, in addition to the
Schwarzschild black hole and Taub-Newman-Unti-Tamburino (Taub-NUT) spacetime \cite{AM53-472,JMP4-915},
another simplest exact vacuum solution is the C-metric \cite{PRD2-1359,AP98-98,PRD67-064001,IJMPD15-335},
which represents an accelerating black hole. In fact, it had already been shown \cite{PRD2-1359,GRG15-535}
that the C-metric solution describes a pair of causally separated black holes which accelerate away from
each other due to the presence of strings or struts that are represented by conical singularities. Later, it was shown
\cite{PRD55-7977} that the C-metric can be derived from the metric of two superposed Schwarzschild black
holes by assuming that the mass and location of one of them approaches infinity in an appropriate way. In
recent years, aspects of the accelerating black holes, including global causal structure \cite{CQG23-6745},
quantum thermal properties \cite{PLA209-6}, holographic heat engines \cite{EPJC78-645,JHEP0219144}, black hole
shadows \cite{PRD103-025005}, holographic complexity \cite{PLB823-136731,PLB838-137691}, and so on, have been
investigated extensively. In particular, thermodynamics of the AdS$_4$ C-metric were figured out first
in Ref. \cite{PRL117-131303} and then well-addressed in Refs. \cite{JHEP0517116,PRD98-104038,JHEP0419096,
PLB796-191,CQG38-145031,CQG38-195024,PRD104-086005,2306.16187}, where the first law of thermodynamics
\cite{PRD7-2333,PRD13-191} and the Bekentein-Smarr mass formula \cite{PRL30-71} as well as the
Christodoulou-Ruffini-type squared-mass formula \cite{PRL25-1596,PRD4-3552} are properly extended to
accelerating, charged, and rotating black holes.

Naturally, the establishment of the above mass formulas is not the only aspect of the investigation of black
hole thermodynamics. Recently, topology has attracted a lot of attention as a mathematical tool to explore
the thermodynamic properties of black holes \cite{PRD105-104003,PRD105-104053,PLB835-137591,PRD107-046013,
PRD107-106009,JHEP0623115,2305.05595,2305.05916,2305.15674,2305.15910,2306.16117,PRD106-064059,PRD107-044026,
PRD107-064015,2212.04341,2302.06201,2304.14988}.\footnote{One can also apply topology to study the light rings
\cite{PRL119-251102,PRL124-181101,PRD102-064039,PRD103-104031,PRD105-024049} and the timelike circular orbits
\cite{PRD107-064006,2301.04786}.} Remarkably, a novel approach proposed in Ref. \cite{PRL129-191101}
has emerged to examine the thermodynamic topological properties of black holes. This approach interprets
black hole solutions as topological thermodynamic defects, establishes topological numbers, and subsequently
categorizes black holes into three distinct classes based on their respective topological numbers. This
groundbreaking methodology has illuminated new facets of our understanding of the fundamental properties
of black holes and gravity. The topological approach outlined in Ref. \cite{PRL129-191101} has gained
widespread acceptance due to its adaptability and simplicity. Consequently, it has been successfully
employed to investigate the topological numbers associated with various well-known black hole solutions
\cite{PRD107-064023,JHEP0123102,PRD107-024024,PRD107-084002,PRD107-084053,2303.06814,2303.13105,2304.02889,
2306.13286,2304.05695,2306.05692,2306.11212,EPJC83-365,2306.02324}. However, the topological number of the
accelerating black holes remains virgin territory, it deserves to be explored deeply, which motivates us to
conduct the present work.

In this paper, we shall investigate the topological number associated with the four-dimensional static
accelerating black hole and its AdS extension, as well as the static charged accelerating black hole and its
AdS extension. This paper aims to fill the gap in the existing literature by examining the influence of the
acceleration parameter on the topological number of black holes, a facet that has been overlooked so far.
The findings of this research will provide valuable insights into the crucial role played by the acceleration
parameter in determining the topological number of static black holes and their AdS counterparts within the
framework of the Einstein-Maxwell gravity theory. We shall witness a constant unity in the difference
of the topological number between the asymptotically flat static accelerating black hole and its corresponding
asymptotically flat static nonaccelerating black hole. Additionally, we shall observe a consistent $-1$ difference
in the topological number between the asymptotically AdS static accelerating black hole and its corresponding
asymptotically AdS static nonaccelerating black hole. We conjecture that they may also be valid for other
accelerating black holes.

The remaining part of this paper is organized as follows. In Sec. \ref{II}, we present a brief review of the
thermodynamic topological approach outlined in Ref. \cite{PRL129-191101}. In Sec. \ref{III}, we first
investigate the topological number of the four-dimensional static accelerating black hole by considering the
simplest static C-metric solution, and then extend it to the case of the static AdS C-metric solution with a nonzero
negative cosmological constant. In Sec. \ref{IV}, we discuss the topological number of the
four-dimensional charged accelerating black hole by considering the Reissner-Nordstr\"om C-metric
(RN C-metric) solution, and then extend it to the RN-AdS C-metric case. Finally, our conclusion and outlook
are given in Sec. \ref{V}.

\section{A brief review of thermodynamic topological approach}\label{II}
In this section, we give a brief review of the novel thermodynamic topological approach. According to Ref.
\cite{PRL129-191101}, we begin by introducing the generalized off shell Helmholtz free energy
\be\label{FE}
\mathcal{F} = M -\frac{S}{\tau}
\ee
for a black hole thermodynamical system with the mass $M$ and the entropy $S$, where $\tau$ is an extra
variable that can be viewed as the inverse temperature of the cavity surrounding the black hole. Only when
$\tau = T^{-1}$, the generalized Helmholtz free energy (\ref{FE}) manifests its on shell characteristics
and converges to the standard Helmholtz free energy $F = M -TS$ of the black hole \cite{PRD98-104038,
PRD15-2752,PRD33-2092,PRD105-084030,PRD106-106015}.

In Ref. \cite{PRL129-191101}, a key vector $\phi$ is defined as
\bea\label{vector}
\phi = \Big(\frac{\p \mathcal{F}}{\p r_{h}}\, , ~ -\cot\Theta\csc\Theta\Big) \, .
\eea
Within the given framework, the parameters are subject to the conditions $0 < r_h < +\infty$ and
$0 \le \Theta \le \pi$. It is important to highlight that the component $\phi^\Theta$ exhibits
divergence at $\Theta = 0$ and $\Theta = \pi$, implying an outward direction of the vector in these
particular scenarios.

A topological current can be established through the utilization of Duan's theory \cite{SS9-1072,
NPB514-705,PRD61-045004} on $\phi$-mapping topological currents in the following manner:
\be\label{jmu}
j^{\mu}=\frac{1}{2\pi}\epsilon^{\mu\nu\rho}\epsilon_{ab}\p_{\nu}n^{a}\p_{\rho}n^{b}\, . \qquad
\mu,\nu,\rho=0,1,2,
\ee
Here, we have $\p_{\nu}= \p/\p x^{\nu}$ and $x^{\nu}=(\tau,~r_h,~\Theta)$. The unit vector
$n$ is formulated as $n = (n^r, n^\Theta)$, where $n^r = \phi^{r_h}/||\phi||$ and $n^\Theta
= \phi^{\Theta}/||\phi||$.  It is evident that the conservation of the aforementioned current
(\ref{jmu}) can be easily demonstrated, leading to $\p_{\mu}j^{\mu} = 0$. Furthermore, it can be
promptly shown that the topological current represents a $\delta$-function of the field configuration
\cite{PRD102-064039,NPB514-705,PRD61-045004}
\be
j^{\mu}=\delta^{2}(\phi)J^{\mu}\Big(\frac{\phi}{x}\Big) \, ,
\ee
where the three-dimensional Jacobian $J^{\mu}(\phi/x)$ fulfills: $\epsilon^{ab}J^{\mu}(\phi/x)
= \epsilon^{\mu\nu\rho}\p_{\nu}\phi^a\p_{\rho}\phi^b$. It becomes evident that the  value of
$j^\mu$ vanishes only when $\phi^a(x_i) = 0$, allowing us to derive the topological number $W$
in the subsequent manner,
\be
W = \int_{\Sigma}j^{0}d^2x = \sum_{i=1}^{N}\beta_{i}\eta_{i} = \sum_{i=1}^{N}w_{i}\, .
\ee
In the given context, $\beta_i$ represents the positive Hopf index, serving as a count for the number
of loops formed by the vector $\phi^a$ within the $\phi$-space as $x^{\mu}$ revolves around the zero
point (ZP) $z_i$. Simultaneously, $\eta_{i}= \mathrm{sign}(J^{0}({\phi}/{x})_{z_i})=\pm 1$ denotes the Brouwer
degree, and $w_{i}$ denotes the winding number associated with the $i$th zero point of $\phi$ enclosed
within the domain $\Sigma$. Furthermore, in the case that two distinct closed curves, denoted as
$\Sigma_1$ and $\Sigma_2$, encompass the identical zero point of $\phi$, it follows that the corresponding
winding number must be equivalent. Conversely, if there exists no zero point of $\phi$ within the enclosed
region, it is imperative that $W = 0$.

It is important to emphasize that the local winding number $w_{i}$ can serve as a valuable tool for
characterizing the local thermodynamic stability. Thermodynamically stable black holes correspond to positive
values of $w_{i}$, while unstable black holes correspond to negative values. On the other hand, the global
topological number $W$ represents the difference between the numbers of thermodynamically stable and unstable
black holes within a classical black hole solution at a fixed temperature \cite{PRL129-191101}. Hence, the
local winding number not only allows for differentiation between different phases of black holes (stable or
unstable) within the same black hole solution at a specific temperature, but it also facilitates the
classification of black hole solutions based on the global topological number. Moreover, based on this
classification, black holes with the same global topological number exhibit similar thermodynamic properties,
even if they belong to different geometric classes.

\section{Static neutral accelerating black holes}\label{III}
In this section, we will investigate the topological number of the four-dimensional static neutral
accelerating black hole by considering the simplest static C-metric solution, and then extend it to
the case of the static AdS C-metric solution with a nonzero negative cosmological constant.

\subsection{C-metric black hole}
An accelerating black hole can be described by the metric \cite{IJMPD15-335,CJP52-1,CQG20-3269}
\be\label{Cmetric}
ds^2 = \frac{1}{\Omega^2}\left\{-f(r)dt^2 +\frac{dr^2}{f(r)} +r^2\left[\frac{d\theta^2}{g(\theta)}
+g(\theta)\sin^2\theta\frac{d\varphi^2}{K^2} \right] \right\} \, , \quad
\ee
where
\bea
f(r) &=& (1 -A^2r^2)\left(1 -\frac{2m}{r} \right) \, , \nn \\
g(\theta) &=& 1 +2mA\cos\theta \, , \quad \Omega = 1 +Ar\cos\theta \, , \nn
\eea
in which $K$ is the conical deficit of the spacetime, $m$ and $A$ are the mass and acceleration
parameters, respectively.

The thermodynamic quantities are \cite{PRD98-104038,JHEP0419096}
\be\ba\label{ThermC-metric}
&M = \frac{m}{K} \, ,  \qquad \mu_{\pm} = \frac{1}{4}\left(1 -\frac{1 \pm 2mA}{K} \right) \, , \\
&T = \frac{m}{2\pi r_h^2} -\frac{(r_h -m)A^2}{2\pi} \, ,  \quad S = \frac{\pi r_h^2}{K(1 -A^2r_h^2)} \, ,
\ea\ee
where $\mu_{\pm}$ are the tensions of the conical deficits on the north and south poles, $r_h$ are the
locations of the event and Cauchy horizons that satisfy the equation: $f(r_h) = 0$.

It is a simple matter to check that the above thermodynamic quantities simultaneously fulfil the first
law and the Bekenstein-Smarr relation
\bea
dM &=& TdS -\lambda_+d\mu_+ -\lambda_-d\mu_- \, , \\
M &=& 2TS \, ,
\eea
where the thermodynamic lengths \cite{JHEP0517116}
\be
\lambda_{\pm} = \frac{r_h}{1 \pm Ar_h} -m
\ee
are conjugate to the tensions $\mu_{\pm}$.

In the subsequent step, we will derive the topological number of the four-dimensional C-metric
black hole. The evaluation of the Helmholtz free energy for this black hole can be carried out by utilizing
the Euclidean action as follows:
\be\label{Euact1}
I_E = \frac{1}{16\pi}\int_M d^4x \sqrt{g}R +\frac{1}{8\pi}\int_{\p{}M} d^3x \sqrt{h}(\cK -\cK_0) \, ,
\ee
where $h$ is the determinant of the induced metric $h_{ij}$, $\cK$ is the extrinsic curvature of the
boundary, and $\cK_0$ is the subtracted one of the massless C-metric solution as the reference background.
The calculation of the Euclidean action integral yields the following result for the Helmholtz free
energy
\be\label{FECmetric}
F = \frac{I_E}{\beta} = \frac{m}{2K} = M -TS \, ,
\ee
where $\beta = 1/T$ being the interval of the time coordinate. Furthermore, the last equality
of Eq. (\ref{FECmetric}) is valid using the results of Eq. (\ref{ThermC-metric}). Therefore, the conical
singularity (the ($\lambda_\pm-\mu_\pm$)-pairs) has no effect on the calculation of the Helmholtz free energy of
the C-metric black hole. It is interesting to make a comparison of the above discussions with our recent works
\cite{EPJC83-365,2306.02324} on the thermodynamic topology of four-dimensional Taub-NUT spacetimes, where the new
conjugate ($\psi-\mathcal{N}$)-pair  is introduced in the expressions of the Taub-NUT spacetimes. Replacing $T$
with $1/\tau$ in Eq. (\ref{FECmetric}) and substituting $m = r_h/2$, thus the generalized off shell Helmholtz free
energy is
\be
\cF = M -\frac{S}{\tau} = \frac{r_h}{2K} -\frac{\pi r_h^2}{(1 -A^2r_h^2)K\tau} \, .
\ee
Using the definition of Eq. (\ref{vector}), the components of the vector $\phi$ can be easily obtained
as follows:
\bea
\phi^{r_h} &=& \frac{1}{2K} -\frac{2\pi r_h}{(A^2r_h^2 -1)^2K\tau} \, , \\
\phi^{\Theta} &=& -\cot\Theta\csc\Theta \, .
\eea
By solving the equation: $\phi^{r_h} = 0$, one can get a curve on the $r_h-\tau$ plane.
For the four-dimensional static accelerating black hole, one can arrive at
\be\label{tauCmetric}
\tau = \frac{4\pi r_h}{(A^2r_h^2 -1)^2} \, .
\ee
Note that Eq. (\ref{tauCmetric}) consistently reduces to the result obtained in the case of the
four-dimensional Schwarzschild black hole \cite{PRL129-191101} when the acceleration parameter $A$
vanishes.

\begin{figure}[h]
\centering
\includegraphics[width=0.4\textwidth]{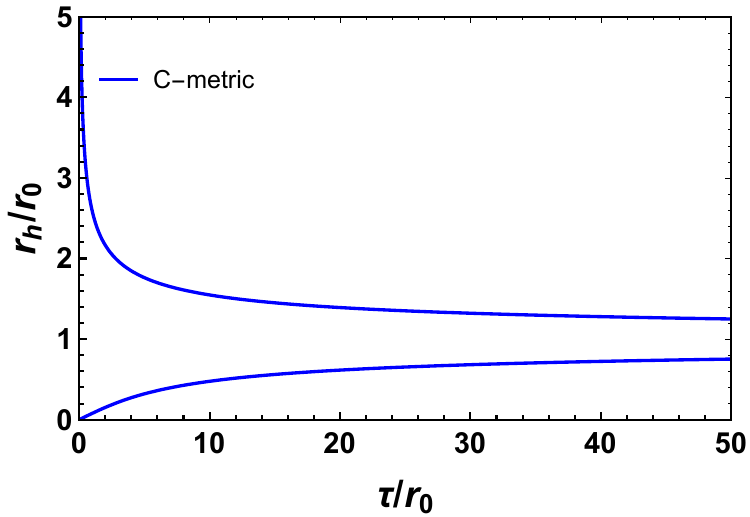}
\caption{Zero points of the vector $\phi^{r_h}$ shown in the $r_h-\tau$ plane
with $Ar_0=1$. There are one thermodynamically stable and one thermodynamically
unstable four-dimensional C-metric black hole for any value of $\tau$.
Obviously, the topological number is: $W = 1 -1 = 0$.
\label{4dCmetric}}
\end{figure}

\begin{figure}[h]
\centering
\includegraphics[width=0.4\textwidth]{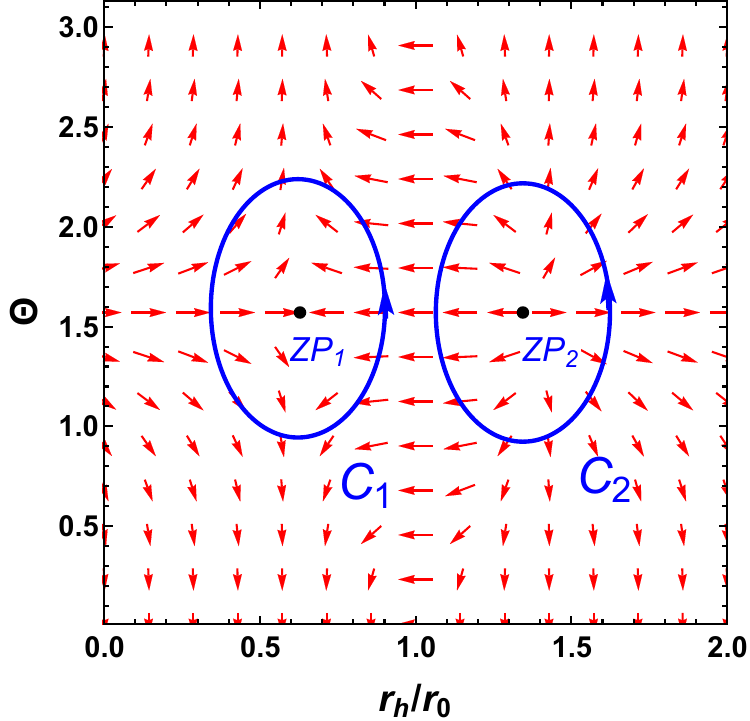}
\caption{The red arrows represent the unit vector field $n$ on a portion of the $r_h-\Theta$
plane for the four-dimensional C-metric black hole with $Ar_0=1$ and
$\tau/r_0 = 20$. The ZPs marked with black dots are at $(r_h/r_0, \Theta)
= (0.62,\pi/2)$, and $(1.39,\pi/2)$, respectively. The blue contours $C_i$ are closed loops
enclosing the zero points.
\label{Cmetric4d}}
\end{figure}

Under the assumption of $Ar_0=1$ for the four-dimensional C-metric black hole (other values of $A$
do not affect the topological number of this black hole), we plot Fig. \ref{4dCmetric} and Fig. \ref{Cmetric4d}
to visualize key aspects. These figures depict the zero points of the component $\phi^{r_h}$ and the behavior of
the unit vector field $n$ on a portion of the $\Theta -r_h$ plane, with $\tau = 20r_0$. Here, $r_0$ corresponds
to an arbitrary length scale determined by the size of a cavity that encloses the static accelerating black hole.
From Fig. \ref{4dCmetric}, one can easily observe that there are one thermodynamically stable and one
thermodynamically unstable four-dimensional C-metric black hole for any value of $\tau$. Therefore,
it is evident that the C-metric black hole exhibits distinct behavior compared to the Schwarzschild black hole
(which always exist one Schwarzschild black hole for any value of $\tau$) \cite{PRL129-191101},
emphasizing the significant impact of the acceleration parameter on the thermodynamical stability of the
static neutral black hole. Consequently, it would be intriguing to explore deeper into the topological properties
of black holes with unusual horizon topologies, such as planar \cite{PRD54-4891}, toroidal \cite{PRD56-3600},
hyperbolic \cite{PRD92-044058}, ultraspinning black holes \cite{PRD89-084007,PRL115-031101,JHEP0114127,PRD103-104020,
PRD101-024057,PRD102-044007,PRD103-044014,JHEP1121031,PRD95-046002,JHEP0118042}, and NUT-charged spacetimes
\cite{PRD100-101501,PRD105-124013,2209.01757,2210.17504,2306.00062}.

In Fig. \ref{Cmetric4d}, the zero points are located at $(r_h/r_0, \Theta) = (0.62,\pi/2)$, and
$(1.39,\pi/2)$, respectively. Consequently, the winding numbers $w_i$ for the blue contours $C_i$
can be interpreted as follows: $w_1 = -1$, $w_2 = 1$, which deviate from those associated with the
Schwarzschild black hole \cite{PRL129-191101}. Regarding the topological global properties, the
topological number $W = 0$ for the four-dimensional C-metric black hole can be readily
observed from Fig. \ref{Cmetric4d}, which also distinguishes it from the topological number of the
Schwarzschild black hole ($W = -1$). Thus, it can be indicated that not only do the C-metric
black hole and the Schwarzschild black hole exhibit clear differences in terms of geometric topology,
but they also belong to different categories from the thermodynamic topological perspective.

\subsection{AdS C-metric black hole}
In this subsection, we will extend the above discussions to the cases of the static neutral AdS
accelerating black hole by considering the four-dimensional AdS C-metric black hole, whose metric
is still given by Eq. (\ref{Cmetric}), but now
\be
f(r) = (1 -A^2r^2)\left(1 -\frac{2m}{r} \right) +\frac{r^2}{l^2} \, , \nn
\ee
in which the AdS radius $l$ is associated with the thermodynamic pressure $P = 3/\big(8\pi{}l^2\big)$ of
the four-dimensional AdS black hole \cite{CPL23-1096,CQG26-195011,PRD84-024037}.

The thermodynamic quantities are \cite{PRD98-104038}
\be\ba\label{ThermAdSC-metric}
&M = \frac{m\alpha}{K} \, ,  \qquad \mu_{\pm} = \frac{1}{4}\left(1 -\frac{1 \pm 2mA}{K} \right) \, , \\
&T = \frac{r_h^3 +ml^2}{2\pi\alpha r_h^2l^2} -\frac{(r_h -m)A^2}{2\pi\alpha} \, ,  \quad
S = \frac{\pi r_h^2}{K(1 -A^2r_h^2)} \, , \\
&V = \frac{4\pi}{3\alpha K}\left[\frac{r_h^3}{\big(1 -A^2r_h^2\big)^2} +mA^2l^4 \right] \, , \quad
P = \frac{3}{8\pi l^2} \, , \\
&\lambda_{\pm} = \frac{1}{\alpha}\left[\frac{r_h}{1 -A^2r_h^2} -m\left(1 \pm \frac{2Al^2}{r_h}\right) \right] \, ,
\ea\ee
where the rescaled factor $\alpha = \sqrt{1 -A^2l^2}$.

It is easy to verify that the above thermodynamic quantities obey the differential first law and integral
Bekenstein-Smarr mass formula simultaneously,
\bea
dM &=& TdS +VdP -\lambda_+d\mu_+ -\lambda_-d\mu_- \, , \\
M &=& 2TS -2VP \, .
\eea

Now, we explore the topological number of the four-dimensional static-neutral accelerating AdS black hole.
In order to get the result of the Helmholtz free energy, one can calculate the Euclidean action
integral \cite{PRD98-104038}
\bea\label{Euact2}
I_E &=& \frac{1}{16\pi}\int_M d^4x \sqrt{g}\Big(R +\frac{6}{l^2}\Big) \nn \\
&&+\frac{1}{8\pi}\int_{\p M} d^3x \sqrt{h}\Big[\cK -\frac{2}{l} -\frac{l}{2}\mathcal{R}(h)\Big] \, ,
\eea
where $\cK$ and $\mathcal{R}(h)$ are the extrinsic curvature and Ricci scalar of the boundary metric
$h_{\mu\nu}$, respectively. To eliminate the divergence, the action encompasses not only the standard
Einstein-Hilbert term but also includes the Gibbons-Hawking boundary term and the corresponding AdS
boundary counterterms \cite{PRD60-104001,PRD60-104026,PRD60-104047,CMP208-413,CMP217-595}.

With the help of the above expressions in Eq. (\ref{ThermAdSC-metric}) and utilizing $m = r_h/2 +r_h^3/[2(1 -\alpha^2r_h^2)l^2]$,
the Helmholtz free energy of the four-dimensional AdS C-metic black hole reads
\bea\label{FEAdSCmetric}
F &=& \frac{I_E}{\beta} = \frac{m\alpha}{2K} -\frac{1}{2\alpha Kl^2}\left[\frac{r_h^3}{(1 -A^2r_h^2)^2} +m\alpha^2l^4 \right] \nn \\
&=& M -TS = \frac{M}{2} -VP  \, ,
\eea
where, similar to the case of the four-dimensional C-metric black hole in the last subsection, the conical singularity also has no effect
on the calculation of the Helmholtz free energy.

Replacing $T$ with $1/\tau$ and substituting $l^2 = 3/(8\pi{}P)$ into Eq. (\ref{FEAdSCmetric}), then the
generalized off shell Helmholtz free energy simply reads
\bea
\cF &=& M -\frac{S}{\tau} \nn \\
&=& \frac{r_h}{24K}\sqrt{16 -\frac{6A^2}{\pi P}}\left(3 +\frac{8\pi Pr_h^2}{1 -A^2r_h^2} \right)
-\frac{\pi r_h^2}{(1 -A^2r_h^2)K\tau} \, . \qquad
\eea
Therefore, the components of the vector $\phi$ are computed as follows:
\bea
\phi^{r_h} &=& \frac{1}{24K(A^2r_h^2 -1)^2}\bigg[-8\sqrt{2\pi}Pr_h^2\sqrt{8\pi -\frac{3A^2}{P}}(A^2r_h^2 -3) \nn \\
&&+3\sqrt{16 -\frac{6A^2}{\pi P}}(A^2r_h^2 -1)^2 \bigg] -\frac{2\pi r_h}{(A^2r_h^2 -1)^2K\tau} \, , \\
\phi^{\Theta} &=& -\cot\Theta\csc\Theta \, ,
\eea
thus one can calculate the zero point of the vector field $\phi^{r_h}$ as
\be
\tau = -\frac{24\pi^{\frac{3}{2}}r_h}{\sqrt{4\pi
-\frac{3A^2}{2P}}\Big[8\pi Pr_h^2(A^2r_h^2 -3) -3\big(A^2r_h^2 -1 \big)^2 \Big]} \, ,
\ee
which consistently reduces to the one obtained in the four-dimensional Schwarzschild-AdS$_4$ black hole case
\cite{PRD107-084002} when the acceleration parameter $A$ is turned off. We point out that the
generation point satisfies the constraint conditions given by
\be
\frac{\p\tau}{\p{}r_h} = 0 \, , \qquad \frac{\p^2\tau}{\p{}r_h^2} > 0 \, ,
\ee
and the annihilation point obeys the constraint conditions as follows
\be
\frac{\p\tau}{\p{}r_h} = 0 \, , \qquad \frac{\p^2\tau}{\p{}r_h^2} < 0 \, .
\ee

\begin{figure}[h]
\centering
\includegraphics[width=0.4\textwidth]{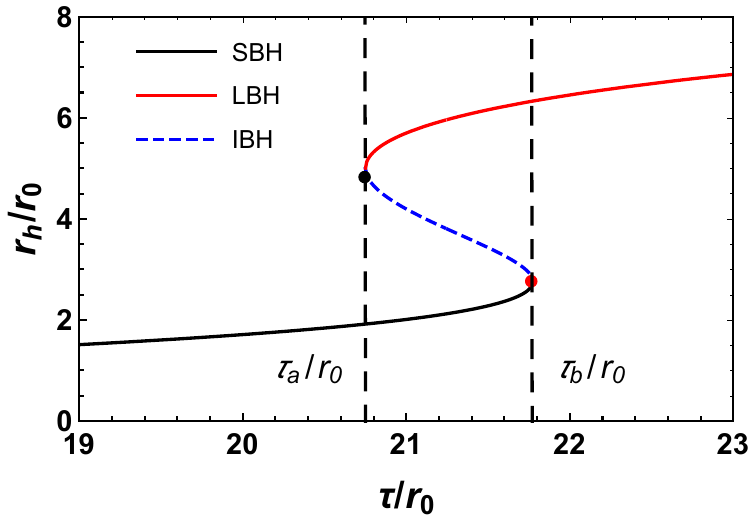}
\caption{Zero points of the vector $\phi^{r_h}$ shown on the $r_h-\tau$ plane
with $Pr_0^2 = 0.01$, and $Ar_0 = 0.2$ for the four-dimensional AdS C-metric
black hole. The red solid, blue dashed, and black solid lines are for the large
black hole (LBH), intermediate black hole (IBH), and small black hole (SBH),
respectively. The annihilation and generation points are represented by red and
black dots, respectively. Clearly, the topological number is: $W = -1 +1 -1 = -1$.
\label{4dAdSCmetric}}
\end{figure}

\begin{figure}
\subfigure[~{The unit vector field for the four-dimensional AdS C-metric black hole with
$\tau/r_0=20$, $Ar_0=0.2$, and $Pr_0^2=0.01$.}]
{\label{AdSCmetric4dtau20}
\includegraphics[width=0.33\textwidth]{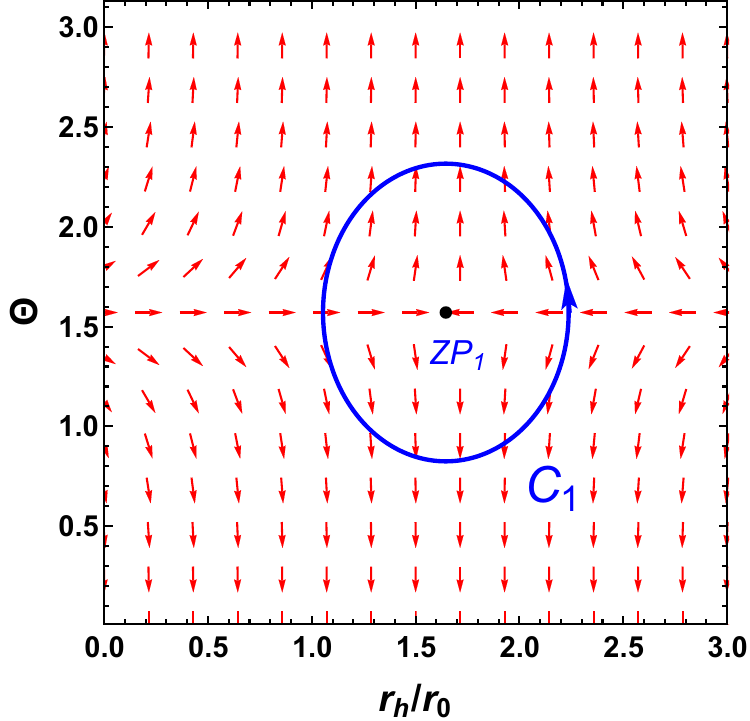}}
\subfigure[~{The unit vector field for the four-dimensional AdS C-metric black hole with
$\tau/r_0=21$, $Ar_0=0.2$, and $Pr_0^2=0.01$.}]
{\label{AdSCmetric4dtau21}
\includegraphics[width=0.33\textwidth]{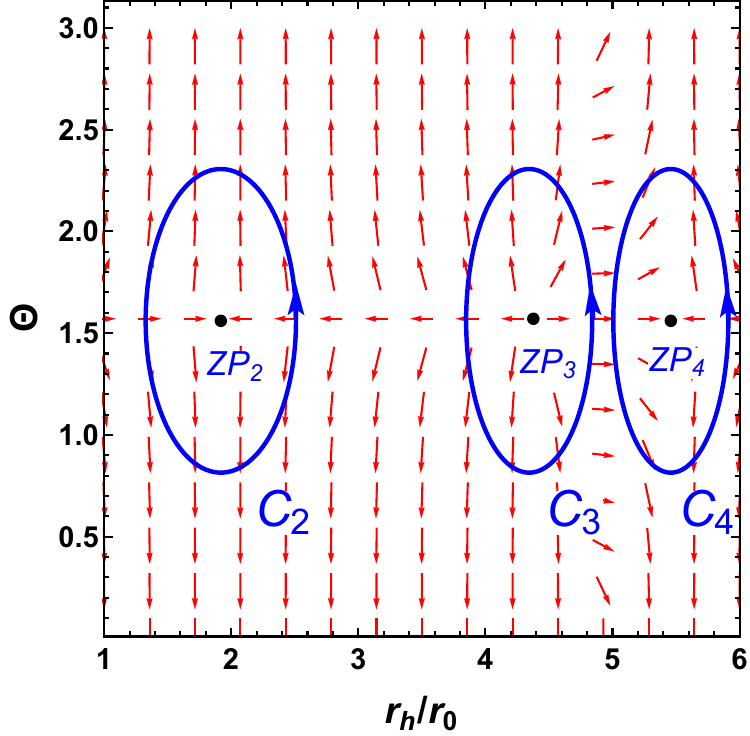}}
\subfigure[~{The unit vector field for the four-dimensional AdS C-metric black hole with
$\tau/r_0=22$, $Ar_0=0.2$, and $Pr_0^2=0.01$.}]
{\label{AdSCmetric4dtau22}
\includegraphics[width=0.33\textwidth]{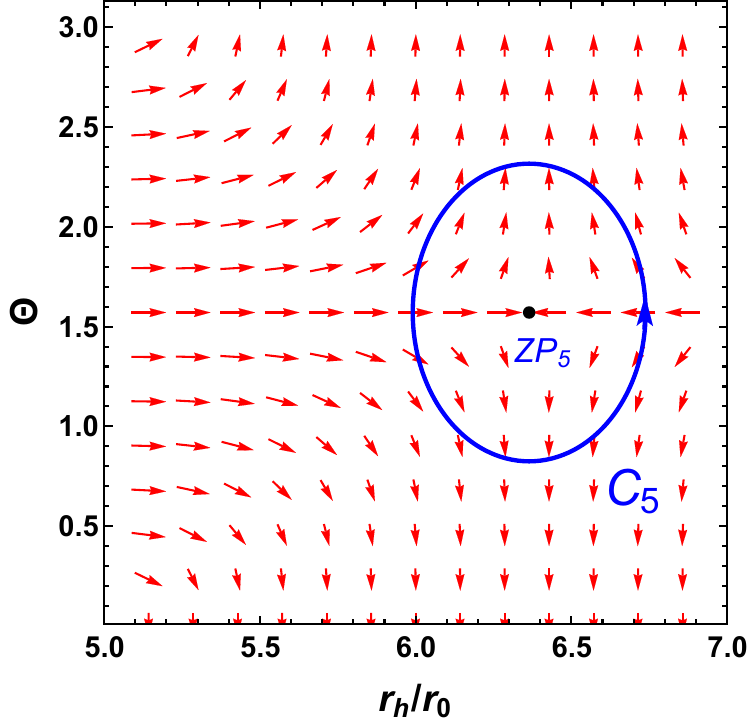}}
\caption{The red arrows represent the unit vector field $n$ on a portion of the $r_h-\Theta$
plane. The ZPs marked with black dots are at $(r_h/r_0,\Theta) = (1.70,\pi/2)$,
$(1.98,\pi/2)$, $(4.37,\pi/2)$, $(5.57,\pi/2)$, $(6.41,\pi/2)$, for ZP$_1$, ZP$_2$, ZP$_3$,
ZP$_4$, and ZP$_5$, respectively. The blue contours $C_i$ are closed loops surrounding the
zero points.
\label{AdSCmetric4d}}
\end{figure}

Considering the pressure as $Pr_0^2 = 0.01$ and the acceleration parameter $Ar_0 = 0.2$ for the
four-dimensional AdS C-metric black hole (other values of $A$ do not influence
its topological number), we illustrate the zero points of $\phi^{r_h}$ in the $r_h-\tau$ plane
in Fig. \ref{4dAdSCmetric}, and the unit vector field $n$ in Fig. \ref{AdSCmetric4d} with
$\tau = 20r_0$, $21r_0$, and $22r_0$, respectively. From Figs. \ref{4dAdSCmetric} and \ref{AdSCmetric4d},
one can observe that for these values of $Pr_0^2$ and $Ar_0$, one generation point and one annihilation
point can be found at $\tau/r_0 = \tau_a/r_0 = 20.75$ and $\tau/r_0 = \tau_b/r_0 = 21.77$, respectively.
It is evident  that there exists a small black hole branch for $\tau < \tau_a$, three distinct black
hole branches for $\tau_a < \tau < \tau_b$, and one large black hole branch for $\tau > \tau_b$.
Computing the winding number $w$ for these three black hole branches, we find that both the small
and large black hole branches have $w = -1$ (thermodynamically unstable), while the intermediate
black hole branch has $w = 1$ (thermodynamically stable). Thus, the AdS$_4$ C-metric black hole
consistently maintains a topological number of $W = -1$, in contrast to the four-dimensional C-metric
black hole discussed in the previous subsection, which possesses a topological number of zero. Therefore,
from a thermodynamic topological standpoint, these aforementioned two black holes represent distinct categories
of black hole solutions, indicating the importance of the cosmological constant in determining the topological
number for the static neutral accelerating black hole. Furthermore, since the topological number of the
Schwarzschild-AdS$_4$ black hole is zero, while that of the AdS$_4$ C-metric black hole is $-1$, it can be
inferred that the acceleration parameter has a remarkable effect on the topological classification of the
four-dimensional static uncharged AdS black hole.

\section{Static charged accelerating black holes}\label{IV}
In this section, we turn to discuss the topological number of the four-dimensional charged accelerating
black hole by considering the RN C-metric solution, and then extend it to the RN-AdS C-metric case with a
nonzero negative cosmological constant.

\subsection{RN C-metric black hole}
A static charged accelerating black hole is represented by the metric and the Abelian gauge potential
\cite{IJMPD15-335}
\bea
ds^2 &=& \frac{1}{\Omega^2}\Bigg\{-f(r)dt^2 +\frac{dr^2}{f(r)} +r^2\bigg[\frac{d\theta^2}{g(\theta)} \nn \\
&&+g(\theta)\sin^2\theta\frac{d\varphi^2}{K^2} \bigg] \Bigg\} \, , \quad \label{RNCmetric} \\
F &=& dB \, , \qquad B = \frac{q}{r}dt \, , \label{Abelian}
\eea
where
\bea
f(r) &=& (1 -A^2r^2)\left(1 -\frac{2m}{r} +\frac{q^2}{r^2} \right) \, , \nn \\
g(\theta) &=& 1 +2mA\cos\theta +q^2A^2\cos^2\theta \, , \quad \Omega = 1 +Ar\cos\theta \, , \nn
\eea
in which $K$ is the conical deficit of the charged accelerating black hole, $m$, $q$, $A$ are the mass,
the electric charge and the acceleration parameters, respectively.

The thermodynamic quantities are \cite{JHEP0419096}
\be\ba\label{ThermRNCmetric}
&M = \frac{m}{\alpha K} \, ,  \qquad \mu_{\pm} = \frac{1}{4}\left(1 -\frac{1 \pm 2mA +q^2A^2}{K} \right) \, , \\
&T = \frac{mr_h -q^2}{2\pi\alpha r_h^3} -\frac{(r_h -m)A^2}{2\pi\alpha} \, ,  \quad Q = \frac{q}{K} \, , \\
&S = \frac{\pi r_h^2}{K(1 -A^2r_h^2)} \, , \qquad \Phi = \frac{q}{\alpha r_h} \, ,
\ea\ee
where the factor $\alpha = \sqrt{1 +q^2A^2}$, $r_h$ are the locations of the event and Cauchy
horizons that obey the horizon equation: $f(r_h) = 0$.

It is easy to verify that the above thermodynamic quantities satisfy the differential first law and the
integral Bekenstein-Smarr relation simultaneously
\bea
dM &=& TdS +\Phi dQ -\lambda_+d\mu_+ -\lambda_-d\mu_- \, , \\
M &=& 2TS +\Phi Q \, ,
\eea
where the thermodynamic lengths
\be
\lambda_{\pm} = \frac{1 \mp Ar_h}{\alpha(1 +q^2A^2)}\left(r_h -2m +\frac{m}{1 \pm Ar_h} \right)
\ee
being conjugate to the tensions $\mu_{\pm}$.

For the four-dimensional RN C-metric black hole, the expression of the Gibbs free energy can be obtained from the
Euclidean action
\be\label{Euact3}
I_E = \frac{1}{16\pi}\int_M d^4x \sqrt{g}\big(R -F^2\big) +\frac{1}{8\pi}\int_{\p{}M} d^3x \sqrt{h}(\cK -\cK_0) \, .
\ee
where $h$ represents the determinant of the induced metric $h_{ij}$, $\cK$ denotes the extrinsic curvature of the
boundary, and $\cK_0$ signifies the subtracted value of the massless C-metric solution used as the reference
background. Thus, the Gibbs free energy reads
\be
G = \frac{I_E}{\beta} = \frac{r_h^2 -q^2}{4\alpha Kr_h} = M -TS -\Phi Q = \frac{M -\Phi Q}{2} \, ,
\ee
where $\beta = T^{-1}$ denotes the interval of the time coordinate, the last two equalities are valid with the
thermodynamic variables in Eq. (\ref{ThermRNCmetric}) as required. In addition, it is easy to see that the conical
singularity also has no effect on the calculation of the Gibbs free energy in this case.

Next, we will investigate the topological number of the four-dimensional static charged accelerating black hole.
We note that the Helmholtz free energy is given by
\be
F = G +\Phi Q = M -TS \, .
\ee
It is very easy to obtain the generalized off shell Helmholtz free energy as
\be
\cF = M -\frac{S}{\tau} = \frac{r_h^2 +q^2}{2\alpha Kr_h} -\frac{\pi r_h^2}{(1 -A^2r_h^2)K\tau} \, .
\ee
Then, the components of the vector $\phi$ are
\bea
\phi^{r_h} &=& \frac{r_h^2 -q^2}{2\alpha Kr_h^2} -\frac{2\pi r_h}{(A^2r_h^2 -1)^2K\tau} \, , \\
\phi^{\Theta} &=& -\cot\Theta\csc\Theta \, .
\eea
Thus, by solving the equation: $\phi^{r_h}$ = 0, one can compute the zero point of the vector field
$\phi$ as
\be\label{tauRNC}
\tau = \frac{4\pi\alpha r_h^3}{(r_h^2 -q^2)(A^2r_h^2 -1)^2} \, .
\ee
We point out that Eq. (\ref{tauRNC}) consistently reduces to the one obtained in the case of the
four-dimensional RN black hole \cite{PRL129-191101} when the accelerating parameter $A$ vanishes.

\begin{figure}[t]
\centering
\includegraphics[width=0.4\textwidth]{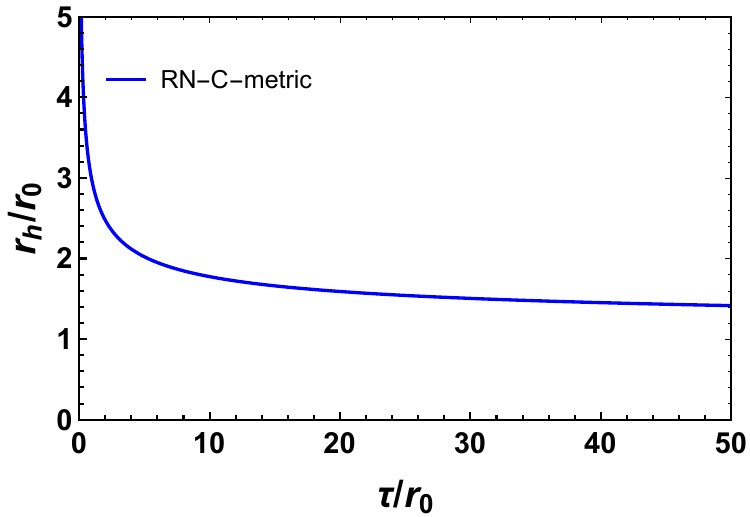}
\caption{Zero points of the vector $\phi^{r_h}$ shown in the $r_h-\tau$ plane with
$Ar_0 = 1$ and $q/r_0 = 1$ for the four-dimensional RN C-metric black hole. There
is one thermodynamically stable four-dimensional RN-C-metric black hole for any
value of $\tau$. Obviously, the topological number is $W = 1$.
\label{4dRNCmetric}}
\end{figure}

\begin{figure}[h]
\centering
\includegraphics[width=0.4\textwidth]{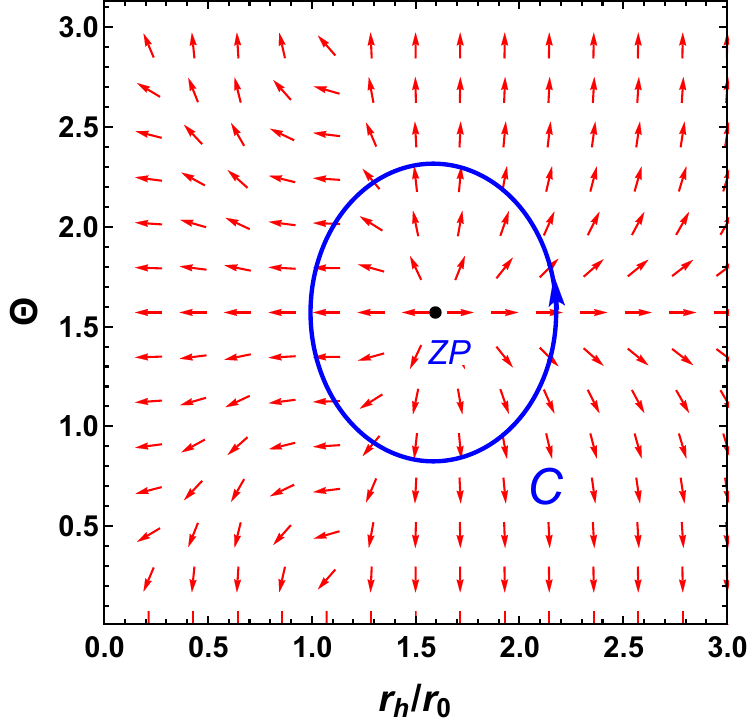}
\caption{The red arrows represent the unit vector field $n$ on a portion of the
$r_h-\Theta$ plane for the four-dimensional RN C-metric solution with $Ar_0 = 1$, $q/r_0 = 1$,
and $\tau/r_0 = 20$. The ZP marked with black dot is at $(r_h/r_0,
\Theta) = (1.59, \pi/2)$. The blue contour $C$ is a closed loop enclosing the
zero point.
\label{RNCmetric4d}}
\end{figure}

For the four-dimensional RN C-metric black hole, we take $Ar_0=1$, $q/r_0=1$, and plot the zero points of the
component $\phi^{r_h}$ in Fig. \ref{4dRNCmetric}, and the unit vector field $n$ with $\tau/r_0=20$ in Fig.
\ref{RNCmetric4d}, respectively. By the way, we also point out that different values (nonzero)
of the acceleration parameter $A$ do not affect the topological number of the four-dimensional RN C-metric black hole.
Obviously, there is only one thermodynamically stable four-dimensional RN C-metric black hole for any value of $\tau$.
Based upon the local property of the zero point, one can obtain the topological number $W = 1$ for the four-dimensional
RN C-metric black hole, which is different from that of the four-dimensional RN black hole ($W = 0$) \cite{PRL129-191101}.
This fact indicates that the acceleration parameter parameter plays an crucial role in determining the topological number
of the four-dimensional static-charged black hole. In addition, compared with the four-dimensional RN black hole which
has a topological number of zero, it can be inferred that the electric charge parameter also has an important effect
on the topological number for the four-dimensional C-metric black hole.

\subsection{RN-AdS C-metric black hole}
In this subsection, we will extend the discussions in the last subsection to the cases of the
static charged accelerating AdS black hole by considering the four-dimensional RN-AdS C-metric black hole,
whose metric and the Abelian gauge potential are still given by Eqs. (\ref{RNCmetric}) and (\ref{Abelian}),
but now
\be
f(r) = (1 -A^2r^2)\left(1 -\frac{2m}{r} +\frac{q^2}{r^2} \right) +\frac{r^2}{l^2} \, . \nn
\ee

The thermodynamic quantities are \cite{JHEP0419096}
\be\ba\label{ThermRNAdSCmetric}
&M = \frac{m[1 -A^2l^2(1 +A^2q^2)]}{\alpha K} \, , \quad S = \frac{\pi r_h^2}{K(1 -A^2r_h^2)} \, , \\
&T = \frac{1}{4\pi\alpha}\bigg[-2A^2r_h\Big(1 -\frac{2m}{r_h} +\frac{q^2}{r_h^2}\Big) \\
&\qquad+2(1 -A^2r_h^2)\Big(\frac{m}{r_h^2} -\frac{q^2}{r_h^3}\Big) +\frac{2r_h}{l^2}\bigg] \, , \\
&Q = \frac{q}{K} \, , \qquad \Phi = \frac{q}{\alpha r_h} \, , \qquad P = \frac{3}{8\pi l^2} \, , \\
&V = \frac{4\pi l^4}{3\alpha Kr_h^5}\Big\{4m^2r_h^2 +mr_h\big[A^2(1 +A^2q^2)r_h^4   \\
&\qquad -4r_h^2 -4q^2\big] +\big(r_h^2 +q^2\big)^2\Big\} \, , \\
&\mu_{\pm} = \frac{1}{4}\left(1 -\frac{1 \pm 2mA +q^2A^2}{K} \right) \, ,
\ea\ee
where the factor $\alpha = \sqrt{1 -A^2l^2(1 +A^2q^2)}\sqrt{1 +A^2q^2}$, and $r_h$ is the largest root of the
horizon equation: $f(r_h) = 0$. Then one can verify that the above thermodynamical quantities completely
satisfy both the the first law and the Bekenstein-Smarr mass formula
\bea
dM &=& TdS +\Phi dQ +VdP -\lambda_+d\mu_+ -\lambda_-d\mu_- \, , \\
M &=& 2TS +\Phi Q -2VP \, ,
\eea
with the thermodynamic lengths
\bea
\lambda_{\pm} &=& -\frac{l^2}{(1 +A^2q^2)\alpha r_h^4}\Big[2m^2r_h(1 -A^2r_h^2) \nn \\
&&+m(Ar_h \mp 1)(Ar_h^3 \pm 2A^2q^2r_h^2 \pm 3r_h^2 +Aq^2r_h \pm q^2) \nn \\
&&\pm r_h(1 +A^2q^2)(A^3q^2r_h^3 \pm r_h^2 -Aq^2r_h \pm q^2)\Big] \, ,
\eea
being conjugate to the tensions $\mu_{\pm}$.

Now, we investigate the Gibbs free energy of the four-dimensional RN-AdS C-metric black hole via the Euclidean
action integral. The expression of the Euclidean action is given as
\bea\label{Euact4}
I_E &=& \frac{1}{16\pi}\int_M d^4x \sqrt{g}\Big(R +\frac{6}{l^2} -F^2\Big) \nn \\
&& +\frac{1}{8\pi}\int_{\p M} d^3x \sqrt{h}\Big[\cK -\frac{2}{l} -\frac{l}{2}\mathcal{R}(h)\Big] \, ,
\eea
where $\cK$ and $\mathcal{R}(h)$ are the extrinsic curvature and Ricci scalar of the boundary metric
$h_{\mu\nu}$, respectively. Along with the standard Einstein-Hilbert term, the action also contains
the Gibbons-Hawking boundary term and the corresponding AdS boundary counterterms in order to eliminate
the divergence. Thus, the Gibbs free energy can simply obtain as
\bea
G &=& \frac{I}{\beta} = \frac{mr_h -q^2}{2\alpha Kr_h} -\bigg\{\frac{2m^2}{\alpha Kr_h^3}
+\frac{(r_h^2 +q^2)^2}{2\alpha Kr_h^5} \nn \\
&&+\frac{m\big[A^2(1 +A^2q^2)r_h^4 -2r_h^2 -2q^2\big]}{\alpha Kr_h^4} \bigg\}l^2 \nn \\
&=& M -TS -\Phi Q = \frac{M -\Phi Q}{2} -VP  \, ,
\eea
where $\beta = T^{-1}$ denotes the interval of the time coordinate, the last two equalities are valid with the
thermodynamic variables in Eq. (\ref{ThermRNAdSCmetric}) as required. Furthermore, it is easy to see that the conical
singularity does not affect the calculation of the Gibbs free energy in this case either.

In order to establish the thermodynamic topological number of the four-dimensional static charged
accelerating AdS black hole, we need to obtain the expression of the generalized off-shell Helmholtz
free energy in advance. The Helmholtz free energy is given by
\be
F = G +\Phi Q = M -TS \, .
\ee
Using the definition of the generalized off shell Helmholtz free energy (\ref{FE}) and
$l^2 = 3/(8\pi{}P)$, one can easily get
\bea
\cF &=& \frac{\sqrt{8\pi -3(A^4q^2 +A^2)P^{-1}}}{4\sqrt{2\pi(1 +A^2q^2)}Kr_h}\bigg[r_h^2 +q^2
+\frac{8\pi Pr_h^4}{3(1-A^2r_h^2)} \bigg] \nn \\
&&-\frac{\pi r_h^2}{(1 -A^2r_h^2)K\tau} \, .
\eea
Thus, the components of the vector $\phi$ are computed as follows:
\bea
\phi^{r_h} &=& \frac{\sqrt{16\pi -6(A^4q^2 +A^2)P^{-1}}}{24\sqrt{\pi(1 +A^2q^2)}(A^2r_h^2 -1)^2Kr_h^2}
\Big[3(r_h^2 -q^2)\big(A^2r_h^2 -1\big)^2 \nn \\
&&-8\pi Pr_h^4(A^2r_h^2 -3)\Big] -\frac{2\pi^{\frac{3}{2}}r_h\sqrt{1
+A^2q^2}}{\sqrt{\pi(1 +A^2q^2)}(A^2r_h^2 -1)^2K\tau} \, , \quad \nn \\
\phi^{\Theta} &=& -\cot\Theta\csc\Theta \, , \nn
\eea
So the zero point of the vector field $\phi$ is
\bea
\tau &=& -\frac{24\pi^{\frac{3}{2}}r_h^3\sqrt{2(1 +A^2q^2)}}{\sqrt{8\pi -3(A^4q^2 +A^2)P^{-1}}}
\Big[8\pi Pr_h^4\big(A^2r_h^2 -3\big) \nn \\
&&+3(q^2 -r_h^2)\big(A^2r_h^2 -1\big)^2 \Big] \, .
\eea

\begin{figure}[t]
\centering
\includegraphics[width=0.4\textwidth]{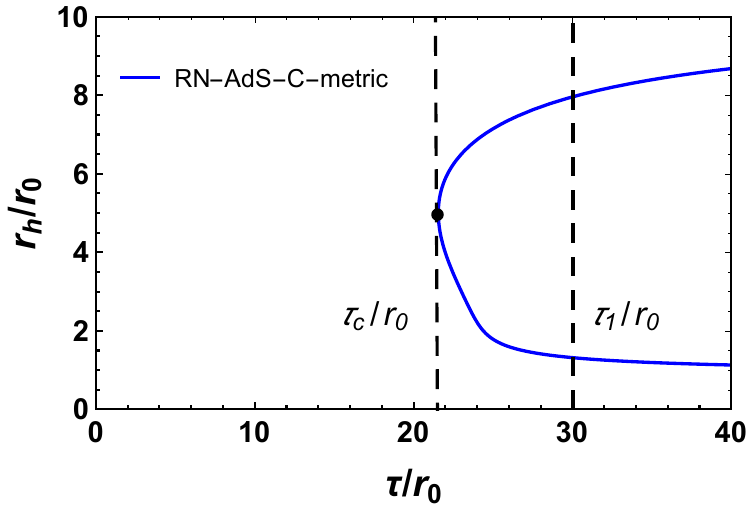}
\caption{Zero points of the vector $\phi^{r_h}$ shown on the $r_h-\tau$ plane with
$q/r_0 = 1$, $Ar_0 = 0.2$ and $Pr_0^2 = 0.01$ for the four-dimensional RN-AdS C-metric
solution. The generation point for this black hole is represented by the black dot
with $\tau_c$. There are two four-dimensional RN-AdS C-metric black holes
when $\tau = \tau_1$. Clearly, the topological number is: $W = -1 +1 = 0$.
\label{4dRNAdSCmetric}}
\end{figure}

\begin{figure}[t]
\centering
\includegraphics[width=0.4\textwidth]{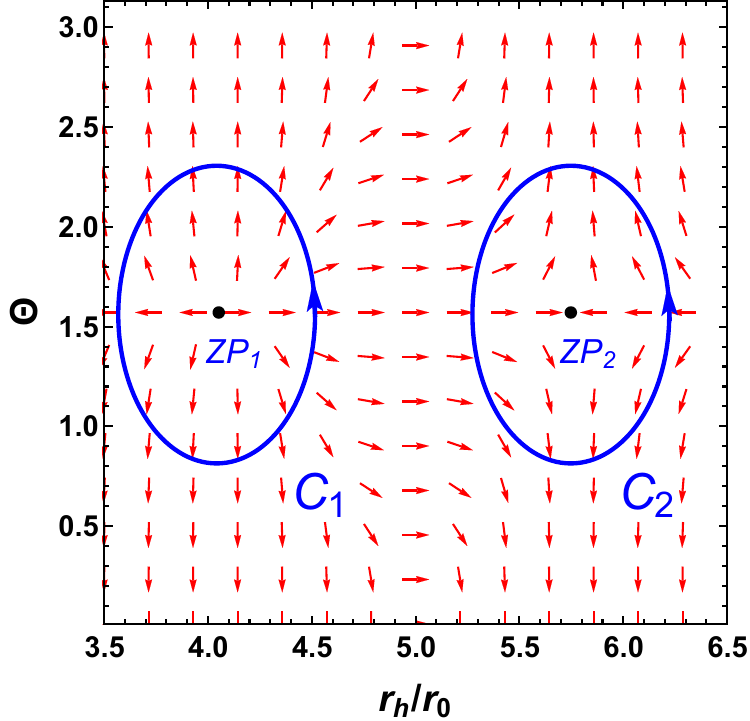}
\caption{The red arrows represent the unit vector field $n$ on a portion of the
$r_h-\Theta$ plane with $q/r_0 = 1$, $Ar_0 = 0.2$, $Pr_0^2 = 0.01$ and $\tau/r_0 = 22$
for the four-dimensional RN-AdS C-metric black hole. The ZPs marked with black dots are
at $(r_h/r_0, \Theta) = (4.11, \pi/2)$, $(5.78, \pi/2)$ for ZP$_1$ and ZP$_2$,
respectively. The blue contours $C_i$ are closed loops surrounding the zero points.
\label{RNAdSCmetric4d}}
\end{figure}

Similar to the procedure adopted before, for the four-dimensional RN-AdS C-metric black hole, we show the zero
points of the component $\phi^{r_h}$ with $q = r_0$, $Ar_0 = 0.2$, and $Pr_0^2 = 0.01$ in Fig. \ref{4dRNAdSCmetric},
and the unit vector field $n$ with $\tau = 22r_0$, $q = r_0$, $Ar_0 = 0.2$, and $Pr_0^2 = 0.01$ in Fig.
\ref{RNAdSCmetric4d}. In addition, we also point out that different values of $A$ have no effect on
the topological number of the four-dimensional RN-AdS C-metric black hole. Note that for these values of $q = r_0$,
$Ar_0 = 0.2$ and $Pr_0^2 = 0.01$, in Fig. \ref{4dRNAdSCmetric}, one generation point can be found at
$\tau/r_0 = \tau_c/r_0 = 21.56$, and there are a thermodynamically unstable four-dimensional RN-AdS C-metric black
hole and a thermodynamically stable four-dimensional RN-AdS C-metric black hole when $\tau = \tau_1$. Based on the
local property of the zero points, we obtain the topological number of the four-dimensional RN-AdS C-metric black
hole is $W = 0$, while that of the RN-AdS black hole is $W = 1$ \cite{PRL129-191101}. Consequently, the introduction
of the acceleration parameter brings about a substantial transformation in the topological number of the four-dimensional
RN-AdS black hole. Moreover, the contrasting topological numbers between the four-dimensional AdS C-metric black hole
($W = -1$) and the four-dimensional RN-AdS C-metric black hole ($W = 0$) underscores the noteworthy impact of the
electric charge parameter on the topological number for the former. Furthermore, the topological number of
$W = 1$ exhibited by the RN C-metric black hole distinguishes it from the RN-AdS C-metric black hole ($W = 0$),
emphasizing the important role played by the cosmological constant in determining the topological number for
the four-dimensional RN C-metric black hole.

\section{Conclusions}\label{V}

\begin{table}[h]
\caption{The topological number $W$, numbers of generation and annihilation points
for static accelerating black holes and their usual nonaccelerating counterparts.}
\resizebox{0.48\textwidth}{!}{
\begin{tabular}{c|c|c|c}
\hline\hline
BH solution & $W$ & Generation point &Annihilation point\\ \hline
C-metric & 0 & 0 & 0\\
Schwarzschild \cite{PRL129-191101} & -1 & 0 & 0 \\\hline
AdS-C-metric & -1 & 1 or 0 & 1 or 0\\
Schwarzschild-AdS \cite{PRD106-064059} & 0 & 0 & 1 \\\hline
RN-C-metric & 1 & 0 & 0\\
RN \cite{PRL129-191101} & 0 & 1 & 0 \\\hline
RN-AdS-C-metric & 0 & 1 & 0 \\
RN-AdS \cite{PRL129-191101} & 1 & 1 or 0 & 1 or 0 \\
\hline\hline
\end{tabular}}
\label{TableI}
\end{table}

The results we found in the current paper are presented in Table \ref{TableI}.
Note that we have also included some known results in the table for comparison
purposes.

In this paper, employing the generalized off shell Helmholtz free energy, we investigate the topological
numbers of the four-dimensional static accelerating black hole and its AdS extension, along with the static
charged accelerating black hole and its AdS extension. We observe that the four-dimensional C-metric black
hole and the AdS$_4$ RN-C-metric black hole fall under the same category of topological classifications,
as evidenced by their same topological number of $W = 0$. On the other hand, the four-dimensional
AdS C-metric black hole and the four-dimensional RN C-metric black hole belong to other two distinct
topological categories, distinguished by their topological numbers of $W = -1$ and $W = 1$, respectively.
By the way, it is worth to noting that different values (nonzero) of the acceleration parameter
do not influence the topological number of the corresponding four-dimensional static accelerating black hole.

Furthermore, it will become apparent that the difference in the topological number between the
asymptotically flat static accelerating black hole and its corresponding asymptotically flat static
nonaccelerating black hole is consistently unity. Moreover, we will take note of the difference in the
topological number between the asymptotically AdS static accelerating black hole and its corresponding
asymptotically AdS static nonaccelerating black hole, which is always $-1$. We conjecture that they
might also hold for other accelerating black holes. However, the conjecture need  to be tested by further
investigating the topological numbers of many other accelerating black holes and their common counterparts.
Through our analysis, we uncover a profound and significant impact of the acceleration parameter on the
topological characteristics of the static black holes.

Additionally, we provide evidence of the crucial role played by the electric charge parameter in determining the topological
number for the static neutral accelerating black holes. What is more, we emphasize the remarkable influence exerted by the
cosmological constant on the topological number of the static accelerating black hole. A most related issue is to extend
the present work to the more general rotating charged accelerating black holes.

\acknowledgments
We are greatly indebted to the two anonymous referees for their constructive comments to improve
the presentation of this work. This work is supported by the National Natural Science Foundation of China
(NSFC) under Grants No. 12205243, No. 12375053, and No. 11675130, by the Sichuan Science and Technology Program under Grant
No. 2023NSFSC1347, and by the Doctoral Research Initiation Project of China West Normal University under Grant
No. 21E028.

\end{document}